\documentclass[structabstract]{aa}
\usepackage{graphicx}
\usepackage{txfonts}
\begin{document}
   \title{Diversity of multiwavelength emission bumps in the GRB\,100219A afterglow}


\author{J. Mao
          \inst{1,2,3,4}
          \and
          D. Malesani\inst{5}
          \and
          P. D'Avanzo\inst{4}
          \and
          S. Covino\inst{4}
          \and
          S. Li\inst{2,3}
          \and
          P. Jakobsson\inst{6}
          \and
          J. M. Bai\inst{2,3}
          }
   \institute{Space Science Division, Korea Astronomy and Space Science Institute, 776, Daedeokdae-ro, Yuseong-gu,
   Daejeon, 305-348, Republic of Korea\\
   \email{jirongmao@mail.ynao.ac.cn}
          \and
Yunnan Astronomical Observatory, Chinese Academy of Sciences,
              Kunming, Yunnan Province, 650011, China\\
         \and
             Key Laboratory for the Structure and Evolution of Celestial Objects, Chinese Academy
             of Sciences, Kunming, Yunnan Province, 650011, China\\
         \and
             INAF-Osservatorio Astronomico di Brera, Via Bianchi 46, I-23807, Merate (LC), Italy \\
             \and
             Dark Cosmology Centre, Niels Bohr Institute, University of Copenhagen, Juliane Maries
             Vej 30, 2100, Copenhagen, Denmark \\
          \and
             Centre for Astrophysics and Cosmology, Science
             Institute, University of Iceland, Dunhagi 5, 107 Reykjavik, Iceland \\
             }

   \date{Received ?, ?, ?; accepted ?, ?, ?}


  \abstract
   {Multi-wavelength observations of gamma-ray burst (GRB) afterglows provide
   important information about the activity of their central engines and their
   environments. In particular, the short timescale variability, such as bumps
   and/or rebrightening features visible in the multi-wavelength light
   curves, is still poorly understood.}
   {We analyze the multi-wavelength observations of the GRB\,100219A afterglow at redshift
   4.7. In particular, we attempt to identify the physical origin of the late
   achromatic flares/bumps detected in the  X-ray and optical bands.}
   {We present ground-based optical photometric data and \textit{Swift} X-ray
   observations on GRB\,100219A. We analyzed the temporal behavior of the X-ray
   and optical light curves, as well as the X-ray spectra.}
   {The early flares in the X-ray and optical light curves peak simultaneously at about 1000 s
   after the burst trigger, while late achromatic bumps in the X-ray and optical bands
   appear at about $2\times 10^4$~s after the burst trigger.
    These are uncommon features in the afterglow phenomenology. Considering the
    temporal and spectral properties, we argue that both optical and X-ray
    emissions come from the same mechanism. The late flares/bumps may be
    produced by late internal shocks from long-lasting activity of the central
    engine. An off-axis origin for a structured jet model is also discussed to
    interpret the bump shapes. The early optical bump
    can be interpreted as the afterglow onset, while the early X-ray flare could
    be caused by the internal activity. GRB\,100219A exploded in a dense
    environment as revealed by the strong attenuation of X-ray emission and the
    optical-to-X-ray spectral energy distribution.}
   {}

   \keywords{gamma rays: bursts -- X-rays: general -- ISM: dust,
   extinction -- shock waves
               }

   \maketitle
%

\section{Introduction}

Gamma-ray bursts (GRBs) are detected by high-energy observational
satellites. Ground-based telescopes can be subsequently alerted and
perform follow-up observations. The accurate positions delivered by
the \textit{Swift} satellite provide an excellent opportunity for
multi-wavelength observations. The analysis of GRB light curves can
provide plenty of information about the central engine and the
surrounding environment. A canonical shape has been identified for
the X-ray light curves (Nousek et al. 2006; Zhang et al. 2006), as
obtained by the \textit{Swift} X-ray Telescope (XRT). In the optical
band, following the alert by the \textit{Swift} Burst Alert
Telescope (BAT), ground-based follow-up observations carried out by
robotic telescopes increased the number of well-sampled optical
light curves significantly (see, e.g., Melandri et al. 2008; Klotz
et al. 2009; Rykoff et al. 2009; Cenko et al. 2009). Apparently,
these light curves present different temporal behaviors. It is hard
to identify a uniform characterization for them. An attempt to
classify the temporal properties has been made by Melandri et al.
(2008). Rykoff et al. (2009) attempted to find commonalities within
a sample of optical light curves, finding that the external forward
shock is a possible origin for the overall optical emission.
Panaitescu et al. (2006) have studied in detail the decay of light
curves both in the X-ray and optical bands from a theoretical point
of view. Chromatic breaks identified by comparing optical and X-ray
light curves indicate that most likely the optical and X-ray
emissions arise from a different origin.

While light curves generically decay in time, in several cases
rebrightenings or bumps are observed in the X-ray or optical bands.
These features call for a more detailed investigation of the physics
of GRB and afterglow. However, the situation is quite complicated.
Most rebrightenings are only observed in the X-ray band. Some early
X-ray bumps, usually called X-ray flares, have no corresponding
optical features (see the statistics by Melandri et al. 2008 and
Rykoff et al. 2009, as well as Uehara et al. 2010 for the individual
cases of GRB\,071112C and GRB\,080506). In contrast, but less
frequently, a rebrightening feature may be seen in the optical but
not in the X-ray band (e.g. GRB\,050721; Antonelli et al. 2006).
Some GRBs with both X-ray and optical bumps shown before 1000 s in
the observer's frame have been identified: GRB\,060418 and
GRB\,060607A (Molinari et al. 2007), GRB\,060904B (Klotz et al.
2008) and GRB\,071031 (Kr\"{u}hler et al. 2009). The X-ray flare and
optical bump of GRB\,060418 have the same peak time. However, it is
likely that the X-ray flare has an internal origin while the optical
bump is the result of external shock onset (see the optical
statistics from Oates et al. 2008). The optical bump and giant X-ray
flare of GRB\,060904B are clearly chromatic. The optical rising and
X-ray flare of GRB\,071031 are not exactly simultaneous, but their
observed correlation suggests a common origin caused by late central
engine activity.

More importantly, it is worth noting that GRBs with late
bumps/rebrightenings shown after $10^3$--$10^4$~s in the observer's
frame are very rare. We mention that the X-ray light curves of
GRB\,050502B, GRB\,050724, GRB\,050904, GRB\,060413, GRB\,060906,
and GRB\,070311 have similar late bump/flare/wiggle features. Some
of them have been identified as late X-ray flares (Curran et al.
2008; Bernardini et al. 2011). It is rare, however, to have
well-sampled, complete optical data at the time corresponding to the
late X-ray bumps (see, e.g., Afonso et al. 2011). A broad, late
optical rebrightening was found by Mundell et al. (2007) for
GRB\,061007, but no corresponding bump was seen in the X-ray band.
The late-observed optical emission of GRB\,050724 is likely related
to the X-ray flare peaking at 41.8~ks, but the evidence for an
optical rebrightening is not conclusive (Malesani et al. 2007).
GRB\,070311 is one case with relatively clear evidence of late bumps
in both the X-ray and optical bands (Guidorzi et al. 2007), with
comparable, although not simultaneous, peak times.
The most interesting case is GRB\,071010A, for which a late
rebrightening feature was identified at 0.6 day after the trigger,
visible simultaneously in both the optical/NIR and X-ray bands
(Covino et al. 2008).

Several mechanisms responsible for the X-ray and optical bumps
presented in GRB light curves have been proposed. Kumar et al.
(2008a, b) suggested a model for the early flares powered by
accretion of fall-back matter from the progenitor star. In general,
early and late X-ray flares may have an internal origin although the
detailed physics is unclear (Chincarini et al. 2010; Bernardini et
al. 2011). An early rising of the optical light curve can be
sometimes caused by the onset of forward shock (Oates et al. 2009),
another discussion was given by Melandri et al. (2010). Some late
optical bumps are interpreted by the external shock model. For
example, the optical rebrightening feature of GRB\,021004 seen at
0.1 days after the burst (Holland et al. 2003) can be explained by
the interaction with a clumpy ambient (Lazzati et al. 2002; but see
also Nakar et al. 2003). An early rising feature around 1000 s and a
later rebrightening behavior after $10^4-10^5$ s can be described by
the standard reverse-forward shock model as well (Zhang et al. 2003,
Fan et al. 2005). However, for the short burst GRB\,060313, the
wiggles shown at about $10^4$ s have been proposed to be the result
of late-time central engine activity (Roming et al. 2006). To
explain the late rebrightening of GRB\,070311 (Guidorzi et al.
2007), refreshed shocks were proposed (Rees \& M\'{e}sz\'{a}ros
1998). The achromatic rebrightening feature of GRB\,071010A may be
produced by an episode of discrete energy injection (Covino et al.
2008).

In this paper, we analyze the X-ray and optical light curves of
GRB\,100219A. We report data taken from the 2.4-m telescope at
Gao-Mei-Gu (GMG) and  from the Nordic Optical Telescope (NOT), with
the former providing most of the data. The optical light curve shows
a clear rebrightening around $2\times10^4$~s after the GRB trigger.
A simultaneous X-ray temporal wiggle/small bump is also detected.
Thus, the late achromatic bumps of GRB\,100219A provide us with an
excellent opportunity to investigate the origin of the bumps/flares
in the different observational bands in more detail.

We describe all observations of GRB\,100219A in Section 2. The GMG
and NOT telescopes have similar mirror size and detectors. Being
located at roughly the same latitude in the Northern hemisphere,
paired together they provide an excellent chance for continuing
observations of GRB afterglows. In Section 3 we compare the optical
light curves with the X-ray data, and we identify achromatic bumps
in the X-ray and optical light curves. We also use the
optical-to-X-ray spectral energy distribution (SED), aiming at
investigating whether the late optical and X-ray bumps come from the
same mechanism. The discussion about the possible physical origin is
presented in Section 4. In Section 5, our conclusions are
summarized.

Throughout the paper, we adopt the convention $F_\nu(t,\nu) \propto
t^{-\alpha}\nu ^{-\beta}$ and cosmological parameters $H_0 = 70$
km~s$^{-1}$~Mpc$^{-1}$, $\Omega_\mathrm{m} = 0.3$ and
$\Omega_\Lambda = 0.7$.

\section{Observations}

\subsection{\textit{Swift} observations}

GRB\,100219A was discovered by \textit{Swift}/BAT on 2010 February
19 at $T_0 = 1$5:15:46 UT (Rowlinson et al. 2010). BAT observations
(Baumgartner et al. 2010) reveal that the prompt light curve has a
roughly triangular shape peak starting at $T_0-10$~s, the peak time
is at about $T_0 + 30$~s, and the burst ends at $T_0 + 50$~s. The
pulse duration is $T_{90}=18.8 \pm 5.0$~s. The BAT time-averaged
spectrum can be fitted by a simple power-law model, the power-law
index being $\Gamma = 1.34\pm 0.25$. The fluence in the 15--150 keV
band is $(3.7\pm 0.6) \times 10^{-7}$~erg~cm$^{-2}$.
\textit{Swift}/XRT began observing the location of GRB\,100219A
178.5~s after the BAT trigger (Rowlinson 2010). \textit{Swift}/UVOT
began observations starting 161 s after the BAT trigger as well
(Holland, Kuin \& Rowlinson 2010; Holland \& Rowlinson 2010),
leading to the detection of a variable object.

\subsection{GMG and NOT optical observations}

\begin{figure}
\vspace{2pt}
\includegraphics[width=\columnwidth]{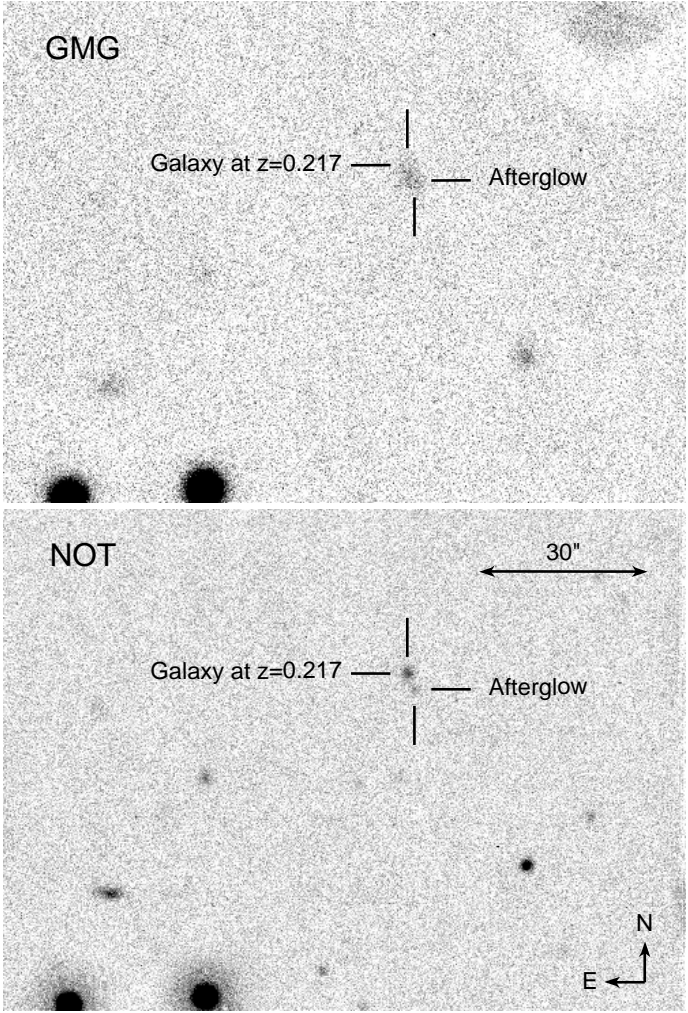}
\caption{Field and optical counterpart of GRB\,100219A as observed
by the NOT (lower panel) and by the GMG (upper panel). The galaxy at
$z=0.217$ is not related to the GRB.}
\end{figure}

Since January 2009, a program of GRB follow-up has been carried out
with the 2.4-m telescope at Lijiang station of Yunnan Astronomical
Observatory, CAS ($\mathrm{longitude}=100\degr 01\arcmin
51\arcsec$~E, $\mathrm{latitude} = 26\degr 42\arcmin 32\arcsec$~N,
$\mathrm{altitude}=3193$~m). The telescope is equipped with standard
Johnson $UBVRI$ filters and a $1340 \times 1300$ pixels CCD with a
pixel scale of $0.214\arcsec/\mathrm{pixel}$. The field of view is
$4\arcmin48\arcsec \times 4\arcmin 40\arcsec$. The GMG telescope
began observing the GRB\,100219A afterglow on 2010 February 19 at
15:30:15 UT, about 15 minutes after the \textit{Swift}/BAT trigger,
taking a sequence of $R$-band observations. The burst location was
observed again on 2010 February 21, starting at 15:14:30 UT, about 2
days after the trigger.

After correcting the raw images with bias and flat fields, we used
the Source Extractor software (SExtractor; Bertin \& Arnouts 1996)
to accurately determine the GRB position. We successfully detected
both the GRB optical afterglow and a nearby faint object, located
3.1\arcsec{} northeast of the afterglow position (Fig. 1, upper
panel). Because the seeing of the GMG observation was poor
($1.8\arcsec$), we were unable to separate the two objects at the
beginning of our observations. As the afterglow was fading, we were
able to clearly identify the two objects in the later images. We
computed the afterglow position using the USNO catalog as reference.
With the $0.15\arcsec$ uncertainty of the USNO catalog (Zacharias
1997), the optical afterglow position of GRB\,100219A is located at
$\mathrm{RA} = 10^{\rm h}~16^{\rm m}~48.54^{\rm s}$, $\mathrm{Dec} =
-12\degr33\arcmin59.5\arcsec$ with an error of $0.34\arcsec$ in the
$R$-band images. This position is fully consistent with the outcome
of the NOT observation (Jakobsson et al. 2010) and the enhanced
\textit{Swift}/XRT position (Evans et al. 2010).

In order to separate the two objects, we first used DAOPHOT (Stetson
1987) to obtain the average point spread function (PSF) in each
image. Then, we used this PSF model to match the afterglow and
nearby faint source. Both sources are pointlike, and with this
procedure we were able to obtain accurate magnitudes of the optical
afterglow.

The 2.5-m NOT ($\mathrm{longitude} = 17\degr 53\arcmin
06.3\arcsec$~W, $\mathrm{latitude} = 28\degr 45\arcmin
26.2\arcsec$~N, $\mathrm{altitude} = 2382$~m) began observing
GRB\,100219A starting on 2010 Feb 20, at 00:11:58.5 UT, about 9~hr
after the trigger. After correcting the raw images for the effect of
CCD flat and bias, we carried out the $R$-band photometry using
DAOPHOT through the selected unsaturated objects in the images to be
consistent with the GMG observation. Since the NOT observations were
carried out under better seeing conditions ($\sim 1.4\arcsec$), the
GRB afterglow and the nearby faint object are resolved.

\begin{figure}
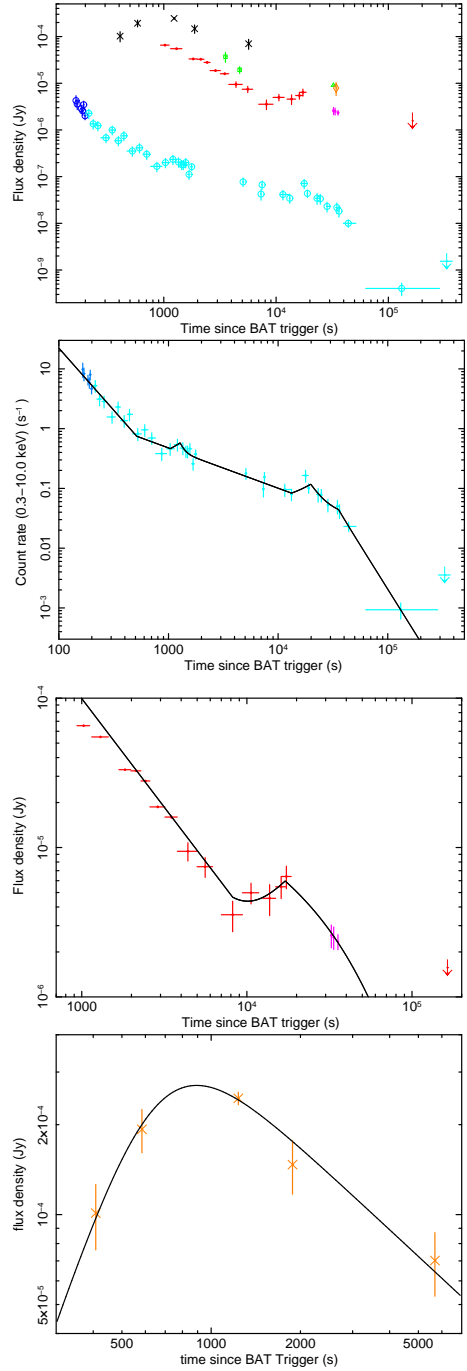

\includegraphics[angle=-90,scale=0.25]{grb100219A_new.ps}
\includegraphics[angle=-90,scale=0.25]{100219Axfit.ps}
\includegraphics[angle=-90,scale=0.25]{100219Aofit_newnew.ps}
\includegraphics[angle=-90,scale=0.25]{Ic_new.ps}
\caption{Panel 1: multi-wavelength light curves of the GRB\,100219A
afterglow. The X-ray light curve at 1~keV is represented by empty
circles (dark blue: WT data; light blue: PC data). The values in the
plot are 10 times lower than the real ones for presentation
purposes. The GMG and NOT $R$-band data are denoted by dots (red:
GMG data; pink: NOT data). The $I$-band datum from the NOT is
indicated by a diamond (orange color). The GAO 150 cm $R_\mathrm{C}$
and $I_\mathrm{C}$ band observations are denoted by empty squares
(green color). The Akeno and Okayama MITSuME optical
$I_\mathrm{C}$-band observations are denoted by crosses (black
color). The GROND $i'$-band observation is marked with a triangle
(green color). Panel 2: the whole X-ray light curve fitting from
about 100 s to $4\times 10^4$ s after the burst trigger and two
X-ray bumps are included in the fitting. Panel 3: the fitting of
optical $R$-band light curve of GMG and NOT observations from about
1000 s to $2\times 10^5$ s after the burst trigger and the late
optical bump is included. Panel 4: the fitting of optical $I_C$ band
light curve of MITSuME observation from about 400 s to 6000 s after
the burst trigger and the early optical bump is clearly shown. The
solid line in each panel is the light curve fitting result.}
\end{figure}

To obtain absolute flux measurements, the standard star field
SA\,105815 from the Landolt (1992) catalog was observed for
photometric calibration. We selected some non-saturated objects in
the images as comparison stars when we performed the photometric
analysis; the magnitude measurements of GMG and NOT are fully
consistent. The final results of the photometry are listed in detail
in Table 1. We present the images from the NOT and GMG in Fig. 1.

\begin{table*}
 \centering
\caption{GMG and NOT observations of GRB\,100219A in 2010. Data are
not corrected for Galactic extinction. }
\begin{tabular}{@{}lcccccccc@{}}
\hline
Telescope & Date & Start time & Time since trigger & Exp. time & Magnitude &Filter & Seeing        \\
          &      & UT         &(s)                 & (s)       &           &       & $(\arcsec)$  \\
\hline
GMG &19 Feb&15:31:22   &   936&  180 ~~~~~&$19.13\pm 0.05$&  $R$   &1.8        \\
GMG &19 Feb&15:34:55   &  1149&  300~~~~~ &$19.32\pm 0.05$&  $R$   &1.8           \\
GMG &19 Feb&15:43:45   &  1679&  300~~~~~ &$19.87\pm 0.08$&  $R$   &1.8           \\
GMG &19 Feb&15:48:48   &  1882&  300~~~~~ &$19.89\pm 0.09$&  $R$   &1.8          \\
GMG &19 Feb&15:53:50   &  2284&  300~~~~~ &$20.06\pm 0.09$&  $R$   &1.8           \\
GMG &19 Feb&15:58:53   &  2587&  600~~~~~ &$20.49\pm 0.13$&  $R$   &1.8         \\
GMG &19 Feb&16:08:54   &  3188&  600~~~~~ &$20.66\pm 0.22$&  $R$   &1.8          \\
GMG &19 Feb&16:18:59   &  3793&  1200~~~~~&$21.23\pm 0.16$&  $R$   &1.8          \\
GMG &19 Feb&16:38:58   &  4992&  1200~~~~~&$21.49\pm 0.18$&  $R$   &1.8            \\
GMG &19 Feb&17:12:29   &  7003&  2400~~~~~&$22.30\pm 0.29$&  $R$   &1.8              \\
GMG &19 Feb&17:52:24   &  9398&  2400~~~~~&$21.93\pm 0.19$&  $R$   &1.8             \\
GMG &19 Feb&18:44:43   & 12537&  2400~~~~~&$22.02\pm 0.29$&  $R$   &1.8              \\
GMG &19 Feb&19:24:42   & 14936&  2400~~~~~&$21.83\pm 0.18$&  $R$   &1.8             \\
GMG &19 Feb&20:04:40   & 17334&  2400~~~~~&$21.66\pm 0.21$&  $R$   &1.8             \\
\hline
GMG &21 Feb&15:14:30   &172784&  2400~~~~~&$>$23.18 (3$\sigma$)& $R$   &1.1               \\
\hline
NOT &20 Feb&00:11:58   & 32172&  600~~~~~ &$22.64\pm 0.21$&  $R$   &1.4               \\
NOT &20 Feb&00:24:27   & 32921&  600~~~~~ &$22.67\pm 0.21$&  $R$   &1.4                 \\
NOT &20 Feb&00:58:44   & 34978&  600~~~~~ &$22.75\pm 0.23$&  $R$   &1.3                    \\
\hline
NOT &20 Feb&00:46:25   & 34239&  300~~~~~ &$21.31\pm 0.41$&  $I$   &1.4                     \\
\hline
\end{tabular}
\end{table*}

\begin{table*}
 \centering
\caption{Optical observations of GRB\,100219A in 2010 taken from the
GCN circulars. GROND started the observation at 00:30 UT. }
\begin{tabular}{@{}lccccccccc@{}}
\hline
Telescope&~~Date  &Middle time &Magnitude &Filter  & Reference \\
\hline
Akeno    &19 Feb&15:22:34    &$18.5\pm 0.3$    &  $I_{\rm C}$ & GCN 10440  \\
\hline
Okayama  &19 Feb&15:25:30    &$17.8\pm 0.2$    &  $I_{\rm C}$ & GCN 10440 \\
Okayama  &19 Feb&15:36:21    &$17.5\pm 0.1$    &  $I_{\rm C}$ & GCN 10440  \\
Okayama  &19 Feb&15:47:07    &$18.1\pm 0.3$    &  $I_{\rm C}$ & GCN 10440  \\
Okayama  &19 Feb&16:50:40    &$18.9\pm 0.3$    &  $I_{\rm C}$ & GCN 10440  \\
\hline
GAO      &19 Feb&16:15:24    &$20.5\pm 0.2$    &  $R_{\rm C}$ & GCN 10452  \\
GAO      &19 Feb&16:34:45    &$19.6\pm 0.3$    &  $I_{\rm C}$ & GCN 10452  \\
\hline
GROND    &20 Feb& about 00:50    & 21.5~~~~        &  $i'$        & GCN 10439 \\
\hline
\end{tabular}
\end{table*}

\subsection{Other optical observations and spectral redshift}

The optical afterglow of GRB\,100219A was also detected by other
ground-based telescopes. Results of their photometric observations
have been reported by Kuroda et al. (2010) at 104~s after the
trigger (MITSuME telescope of the Akeno Observatory and Okayama
Astrophysical Observatory), by Kinugasa et al. (2010) at 0.9~hr
after the trigger (GAO observations), and later by Kr\"uhler et al.
(2010) at 8.7 hr after the trigger (GROND observations).
We collect these results in Table 2.

Spectral measurements were reported by Groot et al. (2010), Cenko et
al. (2010), and de Ugarte Postigo et al. (2010). From the detection
of a series of absorption lines, the redshift was estimated to be $z
= 4.6667\pm 0.0005$ through VLT/X-shooter spectroscopy (de Ugarte
Postigo et al. 2010; Th\"one et al. 2011). The nearby object, with a
redshift of 0.217 (Cenko et al. 2010; Th\"one et al. 2011), is a
galaxy not related to the GRB (as initially suggested by Bloom \&
Nugent 2010).

\section{Results}

From \textit{Swift}/BAT observation, the fluence of GRB\,100219A in
the 15--150 keV band is $(3.7\pm 0.6) \times 10^{-7}~{\rm
erg~cm^{-2}}$. With this fluence, the isotropic energy detected by
BAT is $E_{\rm iso,BAT} \sim 1.4\times 10^{52}~{\rm erg}$. On the
other hand, from the scaling relation of Sakamoto et al. (2009), we
can calculate the peak energy in the observer frame $\log (E_{\rm
peak}/\mathrm{keV}) = 3.258 - 0.829\,\Gamma$ where $\Gamma = 1.34$
is the photon index of the spectrum fitted with a single power-law.
The peak energy $\sim 140$~keV is therefore likely to be above the
spectral range covered by BAT, indicating that the bolometric energy
can be significantly higher. The peak energy in the rest frame is
$E_{\rm peak,rest} = E_{\rm peak} \times (1+z) \sim 800$~keV.
Assuming that GRB\,100219A obeys the $E_{\rm peak,i}-E_{\rm iso}$
relation (``Amati'' relation, Amati et al. 2002, 2006), we predict
that the prompt emission of GRB\,100219A is about $E_{\rm iso}
\approx 7.7 \times 10^{53}$~erg.

\subsection{Light curve features}

In order to analyze the GRB\,100219A multi-wavelength observations
and reveal the involved physics, we also considered the X-ray data
taken by \textit{Swift}. We used the XRT light curve of GRB\,100219A
from the automatic online
repository\footnote{\texttt{http://www.swift.ac.uk/xrt\textunderscore{}curves/00412982/}}.
The flux can be obtained by transforming the count rate with the
conversion factor deduced from the X-ray spectrum. The adopted
procedures are described in detail by Evans et al. (2007, 2009). The
X-ray light curve shows overall the canonical shape (Nousek et al.
2006). However, we note two small bumps/wiggles, at about $10^3$ and
$2\times 10^4$~s.

We converted the GMG and NOT observations from $R$-band magnitude to
flux density. The results of MITSuME, GAO, and GROND are also
reported. The optical light curve shows a brightening as indicated
by the MITSuME $I_\mathrm{C}$-band observation, peaking at about
1000~s. The GMG observations subsequently reveal a clearly fading
phase from about $10^3$ to $10^4$~s. After the decay, the optical
light curve shows a clear rebrightening, peaking at about $2\times
10^4$ s. The multi-wavelength light curves are plotted in Fig. 2. We
note that the early and late bumps shown in the X-ray and optical
light curves are simultaneous.

In order to investigate the temporal behavior of the GRB\,100219A
afterglow in more detail, we performed a fit to the X-ray and
optical light curves. In the fitting process, we focused on the
shapes of the flares/bumps in addition to the general light curves.
The overall X-ray light curve can be fitted by a double-broken
power-law. The decay index of the initial steep phase is $2.07\pm
0.29$. At $635\pm 134$~s, it turns into a relatively flat stage with
a decay index of $0.66 \pm 0.06$. After that, at $(3.5 \pm 1.0)
\times 10^4$~s, the light curve turns again steeply decays with an
index of $2.93\pm 0.29$. The two bumps apparent in the X-ray light
curve were fitted with the same fitting procedure. Chincarini et al.
(2007) fitted the X-ray flares with a Gaussian profile and with the
prompt pulse profile introduced by Norris et al. (2005), showing
that the latter provides a better fit. We adopted the burst model, a
profile with a linear rising, and an exponential
decay\footnote{$F(t) \propto (t-t_{\rm s})/(t_{\rm p}-t{\rm _s})$ if
$t < t_{\rm p}$ and $F(t)\propto \exp(-(t-t_{\rm p})/dt)$ if $t >
t_{\rm p}$, where $t_{\rm s}$ is the start time, $t_{\rm p}$ is the
peak time and $dt$ is the duration. This profile was selected by
Perri et al. (2007) to describe the X-ray flares of GRB\,050730.}.
We obtained for the first flare a peak time of $t_{\rm p} = 1272 \pm
170$~s and for the second flare a peak time of $t_{\rm p} = (1.67
\pm 0.48) \times 10^4$~s. The decay timescales are $157 \pm 237$ s
and $4365 \pm 6948$ s, respectively. The fit has
$\chi^2/\mathrm{dof} = 23.7/33$.

The overall optical light curve obtained by GMG data after 2000~s
can be fitted by a power-law with decay index $1.45\pm 0.04$. After
the time $8203 \pm 2442$~s, the optical light curve shows a
rebrightening feature. With the burst fitting model, the
rebrightening peaks at the time $(1.71\pm 0.21)\times 10^4$~s and
the decay duration is $(2.04\pm 0.67) \times 10^4$~s. The fit has
$\chi^2/\mathrm{dof}=17.0/13$. The F-value of 1.22 and P-value of
0.68 indicate that our fitting is acceptable.

Furthermore, as can be seen in Fig. 2, an early rebrightening is
shown by the $I_\mathrm{C}$-band light curve provided by Kuroda et
al. (2010). We fitted the $I_\mathrm{C}$-band light curve using the
function $F(t) = F_0/[(t/t_{\rm b})^{k\alpha_{\rm r}}+(t/t_{\rm
b})^{k\alpha_{\rm d}}]^{1/k}$ (Molinari et al. 2007), where $t_{\rm
b}$ is break time, $\alpha_{\rm r}$ ($\alpha_{\rm d}$) is the rise
(decay) slope, $k$ is the smoothness parameter, and $F_0$ is
normalization. After the fitting, we obtained $t_{\rm b} = 660\pm
120$ s, $\alpha_{\rm r} = -3.0\pm1.3$, $\alpha_{\rm d} = 0.90\pm
0.20$, $F_0=(4.68\pm0.46)\times 10^{-4}$ Jy, and we fixed the
parameter $k=1$. The fit has $\chi^2/\rm{dof}=5.85/5$. Therefore,
the peak time of the bump is $t_{\rm p} = t_{\rm b}(-\alpha_{\rm
r}/\alpha_{\rm d})^{1/[k(\alpha_{\rm d}-\alpha_{\rm r})]} =
900\pm470$ s. We note that these $I_\mathrm{C}$-band data are
selected from public GCN circular and we do not have any accurate
calibration.

Finally, we plot the fit results in Fig. 2 as well. All errors are
at 1$\sigma$ confidence level\footnote{At the beginning of the
$R$-band light curve, the first two data points (before 1200 s) show
a deviation from the fitting line. These two data points probably
mark the end of the early rebrightnening phase peaking at about 1000
s that are clearly visible in the $I$-band light curve.}. Our fits
quantitatively confirm that the early and late bumps visible in the
X-ray and optical light curves peak simultaneously.

\subsection{Spectral analysis}

Both windowed timing (WT) and photon counting (PC) data of
GRB\,100219A have been collected during the \textit{Swift}/XRT
observation. Here, we consider the PC mode data, which are
simultaneous to our optical data. We downloaded the XRT level 2
cleaned event files and use \texttt{xselect} task to extract the
spectrum. We used the response matrix file from the XRT standard
calibration database and the arf file built by \texttt{xrtmkarf}. We
extracted two spectra for two different time intervals, ranging from
$T_0 + 212$~s to $T_0 + 1814$~s and from $T_0 + 5.0\times 10^3$~s to
$T_0 + 1.8\times 10^5$~s, respectively. Each time interval includes
one X-ray flare. First, we used Swift online
repository\footnote{\texttt{http://www.swift.ac.uk/xrt\textunderscore{}spectra/00412982/}}
(Evans et al. 2007, 2009) to fit the spectra using an absorbed
power-law. We found a spectral index $\beta = 0.69^{+0.17}_{-0.12}$
and an intrinsic column density upper limit $N_{\rm H} \sim 1.9
\times 10^{22}$~cm$^{-2}$ for the first spectrum (reduced $\chi
^2/\rm{dof} = 233.3/254$), and $\beta = 0.86^{+0.16}_{-0.26}$ and
$N_{\rm H} = 7.2^{+4.3}_{-4.7} \times 10^{22}$~cm$^{-2}$ for the
second spectrum (reduced $\chi^2/\rm{dof} = 166.0/186$). The
Galactic column density is $N_{\rm H}=6.5\times 10^{20}$~cm$^{-2}$.

To more thoroughly investigate the potentially large amount of
absorption, we attempted another spectral fit. We fitted the same
spectra as above with a broken power-law. For the first spectrum, we
obtained $\beta_1 = -1.63 \pm 1.79$, $\beta_2 = 0.53 \pm 0.06$ with
the peak energy $E_{\rm p} = 0.60 \pm 0.11$~keV. For the second
spectrum, we obtained $\beta_1 = -0.58 \pm 2.05$, $\beta_2 = 0.55
\pm 0.18$, and $E_{\rm p} = 0.90 \pm 0.44$ keV. The low-energy
spectral indices are poorly determined. The two fits have a reduced
$\chi^2/\rm{dof}$ of $35.2/37$ and $12.9/9$, respectively. All the
spectral parameter errors are at 1$\sigma$ level.

\begin{figure}
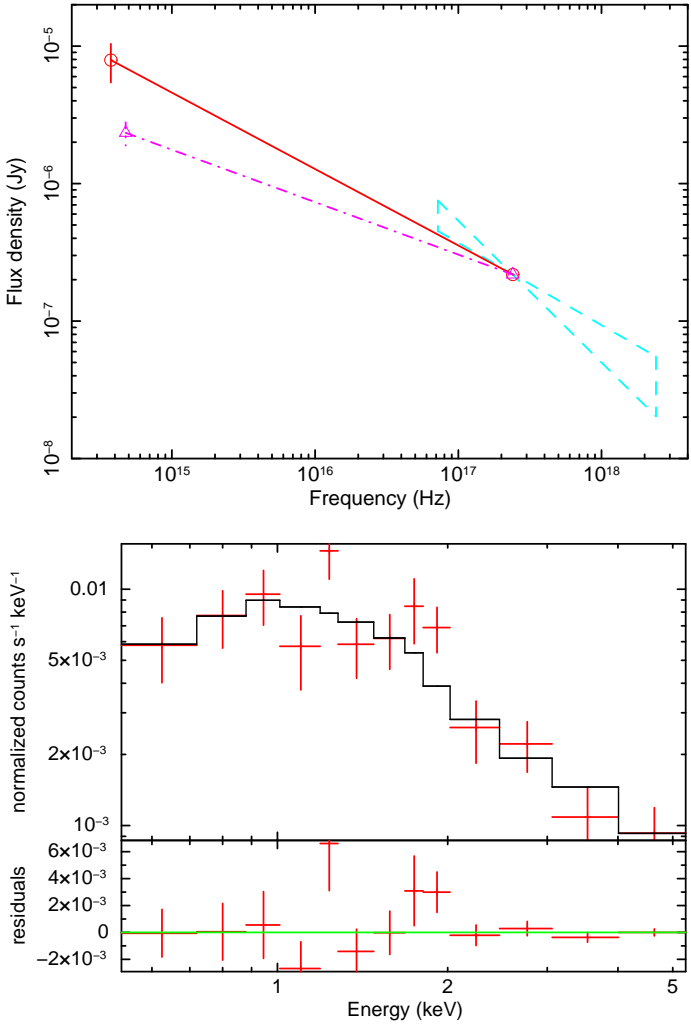

\includegraphics[angle=-90,width=\columnwidth]{ox100219A_newnew_re.ps}
\includegraphics[angle=-90,width=\columnwidth]{bkn_2.ps}
\caption{Upper panel: the optical-to-X-ray spectral energy
distribution of the afterglow of GRB\,100219A is shown in the upper
panel. The X-ray point corresponds to 1~keV. The values of
$\beta_{\rm OX}$ relative to the $I$-band (unaffected by the
Ly$\alpha$ blanketing) and $R$-band (affected by the Ly$\alpha$
blanketing) are 0.56 ($t = 34$~ks) and 0.41 ($t = 35$~ks), marked
with a solid line and a dash-dotted line, respectively. The Galactic
correction $E(B-V)=0.08$ is used. Lower panel: fitting of the late
X-ray spectrum with a broken power-law. The time interval is between
$5.0\times 10^3$ s and $1.8\times 10^5$ s.}
\end{figure}

Using the optical and X-ray fluxes, we also constructed the
optical-to-X-ray SEDs shown in \textbf{Fig. 3}. The X-ray flux
density was computed at 1~keV. Using the $R$-band measurements, we
obtained $\beta_{\rm OX}$ values of 0.27, 0.42, 0.35, and 0.41 at 8,
13, 17, and 35 ks after the burst trigger, respectively. These
values are well below the limit for the definition of dark bursts
first proposed by Jakobsson et al. (2004a). We note that at redshift
$z = 4.7$, the $R$-band flux is strongly suppressed by the
Ly$\alpha$ absorption. This is likely the main reason for the low
value of $\beta_{\rm OX}$. The $I$-band flux is not affected by
Ly$\alpha$ blanketing, and using this flux we obtained $\beta_{\rm
OX} \sim 0.56$ around $3.4\times 10^4$ s.

We compared the $\beta_{\rm OX}$ values with the X-ray spectral
indices. The X-ray fits using a broken power-law are not consistent
with the optical-to-X-ray SED. On the contrary, the X-ray spectral
shape fitted with an absorbed power-law can be roughly extrapolated
to the $I$-band flux level, indicating that the simple power-law
model could fit both the X-ray and optical spectra. Therefore, the
multi-wavelength data seem to require the presence of strong excess
absorption in the X-ray spectra. We note that the $I$-band datum
lies marginally below the extrapolation of the X-ray spectrum. This
can be taken as an indication of moderate dust extinction associated
with the absorbing matter that affects the optical (rest-frame UV)
data. The SED results also suggest that the emission of GRB\,100219A
in both the X-ray and optical bands is coming from the same
mechanism.

\section{Discussion}

As we have listed in Section 1, there are very few cases reported so
far of late flares/bumps observed to peak simultaneously in the
optical and X-ray bands (e.g. Covino et al. 2008 for GRB\,071010A).
The rebrightening features in the optical and X-ray bands of
GRB\,071010A were suggested to originate from a discrete episode of
energy injection. We presented the case of GRB\,100219A, which
provides another example of achromatic bumps.

If we assume that the early optical bump peaking at $895$ s shown in
$I_\mathrm{C}$ band light curve is originally from the onset of the
forward shock (Oates et al. 2009), the Lorentz factor of the burst
shock can be given by $\Gamma(t_{\rm peak})=160[E_{\rm
iso,53}(1+z)^3/n_0t^3_{\rm p,2}]^{1/8}$ from the estimation of
Molinari et al. (2007), where $E_{\rm iso,53}=E_{\rm
iso}/10^{53}~\rm{erg}$ and $E_{\rm{iso}}$ is the isotropic energy
released in gamma band, $t_{\rm p,2}=t_{\rm p}/100~\rm{s}$. With the
density of the surrounding medium $n = 10~\rm{cm^{-3}}$, we obtained
the initial bulk Lorentz factor $\Gamma\approx 260$, which should be
twice of $\Gamma(t_{\rm peak})$. This number is between the value
found by Covino et al. (2008) for GRB\,071010A and the value
measured by Molinari et al. (2007) for GRB\,060418 and GRB\,060607A.
With the value of $E_{\rm{iso}}$ estimated in Section 3, we found
that GRB 100219A has a slight bias to the $\Gamma-E_{\rm iso}$
relation derived by Liang et al. (2010) at the 2$\sigma$ confidence
level. Within the errors, the early bumps shown in the $I_{\rm
C}$-band and X-ray light curves have the same peak time. However,
there are significant differences in the two bumps, e.g., the X-ray
flare has a shorter duration and is sharper than the $I_{\rm
C}$-band bump, suggesting a different origin of these two bumps. The
early X-ray bump may be caused by internal activity, as was that of
GRB\,060418 (Molinari et al. 2007). Furthermore, the smooth decay
($\alpha_{\rm d}=0.9 \pm 0.2$) of the $I_{\rm C}$-band bump is
inconsistent with the rapid-decay prediction ($\alpha_{\rm d}=2.0$)
for the reverse shock emission (Kobayashi \& Zhang 2007). Finally,
we remark that the fitting of the early optical bump cannot be
accurate, because the $I_{\rm C}$-band light curve is poorly sampled
and we do not have enough data points with accurate calibration. In
the following we focus on the late achromatic bumps shown in the
X-ray and optical light curves.

We can calculate the ratio between the duration $\Delta t$ and the
peak time $t_{\rm p}$ of each bump from our light curve fits to
identify the origin of the simultaneous bumps shown in the
GRB\,100219A afterglow. We converted the time duration as computed
in our parametrization to the full-width-at-half-maximum (FWHM) of
the bump. We obtained $(\Delta t_{\rm FWHM}/t_{\rm p})_{\rm X,1} =
0.17 \pm 0.26$ for the first X-ray bump, $(\Delta t_{\rm
FWHM}/t_{\rm p})_{\rm X,2} = 0.66 \pm 0.57$ for the second X-ray
bump and $(\Delta t_{\rm FWHM}/t_{\rm p})_{\rm O}=1.46 \pm 0.58$ for
the late optical bump. These results can be compared with the
predictions by Lazzati \& Perna et al. (2007). As illustrated in
Fig. 2 of their paper, external shock models predict $\Delta t_{\rm
FWHM}/t_{\rm p} \ge 2$. It seems that our X-ray and optical
rebrightening bumps do not favor an external origin. Instead, within
the internal framework, if the flare activity is caused by a freely
expanding flow during the prompt phase, the prediction $\Delta
t_{\rm FWHM} / t_{\rm p} \ge 0.25$ agrees with the observed late
X-ray and optical bumps of GRB\,100219A. The shorter $\Delta t_{\rm
FWHM}/t_{\rm p}$ of the first early and sharp X-ray flare at 1000 s
indicates that the central engine was active for a time
significantly longer than the prompt duration $T_{90}$. This
ejection mechanism has also been proposed by Ghisellini et al.
(2007). Therefore, we suggest that the bumps/flares of GRB\,100219A
have an internal origin from a long-lasting activity of the central
engine.

Another model discussed in the literature to interpret
rebrightenings is the off-axis jet. The bulk Lorenz factor and the
energy per solid angle have in this case some dependency as a
function of the off-axis angle (M\'{e}sz\'{a}ros et al. 1998).
Therefore, the presence of an off-axis jet may modify the temporal
power-law decay and produce a smooth bump (Zhang \& M\'{e}sz\'{a}ros
2002; Kumar \& Granot 2003). In the special case of GRB 100219A, we
considered the possibility proposed by De Pasquale et al. (2009): a
narrow inner jet is responsible for the X-ray emission while the
wide outside jet is responsible for the optical emission. Therefore,
we speculate that the optical bump has much less off-axis effect
than the X-ray bump. Consequently, the optical bump is wider and the
inner X-ray flare is narrower. Unfortunately, in our case, we do not
observe a clear jet-break either in the X-ray or the optical light
curves. This lack of jet-break observation prevents any further
analysis.

Finally, from the X-ray analysis, we suggest that the surrounding
medium of GRB\,100219A might be relatively dense. The neutral
hydrogen column density $1.1\times 10^{22}$~cm$^{-2}$ measured in
the time-averaged X-ray spectrum at $z\sim 4.7$ is higher than the
average value $7.9\times 10^{21}$~cm$^{-2}$ given by the statistics
of Campana et al. (2010). Zheng et al. (2009) and Campana et al.
(2010) have suggested that the GRB X-ray absorption is intrinsic in
general. GRB\,100219A shows a ``dark'' property according to the
criterion $\beta_{\rm OX} \le 0.5$ introduced by Jakobsson et al.
(2004a). However, the low $\beta_{\rm OX}$ value in the $R$-band is
caused by the Ly$\alpha$ blanketing because this burst is at
redshift 4.7. Considering the $I$-band observation, which is not
affected by the Ly$\alpha$ forest, however, $\beta_{\rm OX} = 0.56$,
which is above the criterion by Jakobsson et al. (2004a). By
examining the broad-band SED, we can see that the $I$-band data
point lies (slightly) below the extrapolation of the X-ray spectrum.
This suggests that some flux suppression could be present in the
optical, likely caused by dust, although its significance is not
high. The high metal column density measured in the X-ray spectrum
would suggest a far higher extinction assuming the Galactic
dust-to-metal ratio (Mao 2010). A similar mismatch has been noted by
many authors and is a general feature of GRB afterglows (e.g.,
Galama \& Wijers 2001; Zafar et al. 2011).

From the above analysis, we caution that (1) the X-ray flare sample
selected by Chincarini et al. (2010) is prominently constituted by
bright flares, with a peak count rate above 1 count~s$^{-1}$, while
in our X-ray light curve, the peaks of the two bumps are lower than
this value; this faint feature is therefore hard to model
accurately; (2) for the discussion of the late optical bump, we have
no data points around the peak time, so that we cannot fully
characterize the bump profile; (3) the lack of a jet-break detection
prevents us from carrying out a more detailed analysis; (4) as this
burst occurred at redshift 4.7, in the optical band, we have only
$I$-band observations that are not affected by the Ly$\alpha$
absorption, consequently, the optical-to-X-ray SED cannot be
constructed accurately; (5) although the temporal variabilities can
be related to the central engine and jet structure (see Ioka et al.
2005 for a general description and one application to GRB 080210 by
De Cia et al. 2011), the explanation of small flares/bumps shown in
the GRB light curves is not universal. For example, the optical
wiggles of GRB\,011211 may be the result of spherically asymmetric
density or energy variations (Jakobsson et al. 2004b).

\section{Conclusions}

We have presented the multi-wavelength observations of GRB\,100219A
using the data from \textit{Swift} and ground-based telescopes. The
early (1000 s) and late ($2\times 10^4$ s) achromatic bumps visible
in the X-ray and optical light curves are comprehensively discussed.
In our analysis, we speculate that the early optical $I_{\rm
C}$-band bump is probably caused by the afterglow onset while the
X-ray one might be caused by the internal activity. The late X-ray
and optical bumps may be produced by the internal shell ejection
from the long-active central engine. The jet structure could also
play an important role for the achromatic property of the bumps.
Moreover, the medium surrounding this GRB might be dense.

At present, many theoretical models have been proposed successfully
to solve the problems of GRB energy release and to explain some
major observational phenomena (such as prompt emission, early X-ray
light curve and X-ray flares). However, it is still difficult to
simply apply these models to specify the late achromatic bumps
discussed in this paper. From this point of view, the theoretical
explanations have to concentrate on more subtle observational
features in the future.

\begin{acknowledgements}
We thank the referee, Elena Pian, for the constructive suggestions
and a detailed review. This work made use of XRT data supplied by
the UK \textit{Swift} Science Data Centre at the University of
Leicester. We thank S. Campana for the discussion of X-ray spectrum.
J. Bai, S. Li and J. Mao are financially supported by Natural
Science Foundation of China (NSFC, Grant 10973034) and the 973
Program (Grant 2009CB824800). The Dark Cosmology Centre is funded by
the Danish National Research Foundation.

\end{acknowledgements}

%
%
%
%
%
%
%
%
%

\end{document}